\documentclass[twocolumn,prl]{revtex4}

\ifx\pdfoutput\undefined
  \usepackage[dvips]{graphicx}
  \usepackage[dvips]{color}
\else
  \usepackage[pdftex]{graphicx}
  \usepackage[pdftex]{color}
\fi

\usepackage{dcolumn}
\usepackage{amsmath}
\usepackage{latexsym}
\usepackage{units}
\usepackage{ulem}

\begin{document}

\title[Short Title]{Pumping properties of the hybrid
single-electron transistor in dissipative environment}
\author{S.~V.~Lotkhov}
\email[Electronic mail:~]{Sergey.Lotkhov@ptb.de}
\affiliation{Physikalisch-Technische Bundesanstalt, Bundesallee
100, 38116 Braunschweig, Germany}%
\author{S.~Kafanov}
\author{J.~P.~Pekola}
\affiliation{Low Temperature Laboratory, Helsinki University of
Technology, P.O. Box 3500, 02015 TKK, Finland}%
\author{A.~Kemppinen}
\affiliation{Centre for Metrology and Accreditation (MIKES), P.O.
Box 9, FIN-02151 Espoo, Finland}%
\author{A.~B.~Zorin}
\affiliation{Physikalisch-Technische Bundesanstalt, Bundesallee
100, 38116 Braunschweig, Germany}%

\date{\today}

\begin{abstract}
Pumping characteristics were studied of the hybrid
normal-metal/superconductor single-electron transistor embedded in a
high-ohmic environment. Two $\unit[3]{\mu m}$-long microstrip
resistors of CrO$_{\text x}$ with a sum resistance $R \approx
\unit[80]{k\Omega}$ were placed adjacent to this hybrid device.
Substantial improvement of pumping and reduction of the
subgap leakage were observed in the low-MHz range. At higher
frequencies $\unit[0.1-1]{GHz}$, a slowdown of tunneling due to the
enhanced damping and electron heating negatively affected the pumping, as
compared to the reference bare devices.
\end{abstract}

\maketitle

One of the paramount applications of metallic single electron
tunneling (SET) devices has been generation and detection of very
low quantized currents, $I \sim \unit[1]{pA}$, for the purposes of
charge metrology (see, e.g., Ref.~\cite{Keller}). An important role
plays the fundamental trade-off between the accuracy,
usually requiring opaque tunnel barriers, and the pumping
rates. Sophisticated algorithms are applied to operate the
multi-junction and multi-gate SET pumps at a high accuracy level.
High-frequency pumping (for example, at $f \sim \unit[1]{GHz}$, $I =
ef \sim \unit[160]{pA}$), or driving several pumps in parallel, in
order to produce nA-currents, is theoretically
possible~\cite{AccuracyOfThePump}, but technologically very
challenging.

Recently, single-gate pumping has been demonstrated for the very
simple SET structures, consisting of two ultrasmall Al/AlO$_{\text
{x}}$/Cu Superconductor--Insulator--Normal metal (SIN) contacts,
arranged either as NISIN~\cite{Pekola-NaturePhysics} or
SINIS~\cite{Kemppinen} transistors. The pumping mechanism reproduces
qualitatively the hold-and-pass strategy, known for a four-junction
SET turnstile~\cite{Geerligs}. The mechanism is based on the charge
hysteresis, arising due to the gap $\Delta$ in the energy spectrum
of the superconductor of the SIN junctions.

The pumping accuracy has been analyzed in detail in
Refs.~\cite{Pekola-NaturePhysics,PekolaAverin}. In particular, it
was shown that for the SINIS type devices with a high charging
energy, $E_{\text{C}} \equiv e^2/2C_{\Sigma} > \Delta$ ($C_{\Sigma}$
is the total capacitance of the transistor island), the lowest-order
quantum leakage mechanism, a so-called Cooper-pair---electron (CPE)
cotunneling, involves a coherent tunneling of three particles. This
is an advantage of the hybrid devices, if compared, for example, to
a 3-junction normal-state pump, subjected to two-electron
cotunneling~\cite{Pump}. For realistic SINIS transistors with
$E_{\text {C}}/\Delta \ge 2$ and a rectangular gate drive, the
metrological accuracy of $10^{-8}$ is predicted for the currents
$\sim \unit[10]{pA}$ ~\cite{PekolaAverin,Kemppinen2009}. The device
simplicity opens a possibility of on-chip integration towards higher
currents.

\begin{figure}[b]
\centering%
\includegraphics[width=\columnwidth]{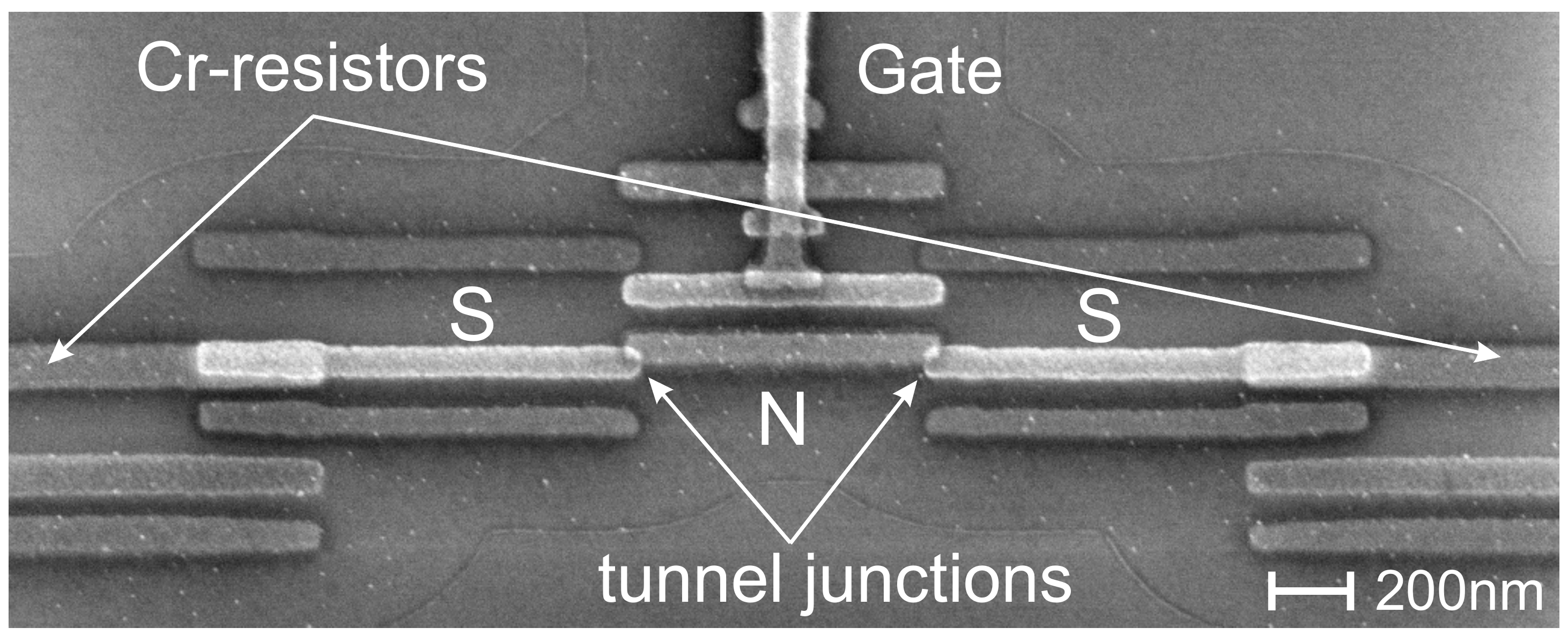}
\caption{Scanning electron micrograph of an R-SINIS device,
fabricated through the triple-replica deposition of CrO$_{\text x}$,
Al and Cr (from top to bottom).}
\label{Fig1}
\end{figure}

In this Letter, we address an important modification of the hybrid
devices due to including dissipative environment, realized as
high-ohmic on-chip microresistors. Our interest is motivated, on one
side, by successful experiments on cotunneling suppression in the
normal-state SET circuits~\cite{R-pump}. On the qualitative level,
similar improvements of the fundamental accuracy are expected
for the hybrid devices as well. On the other hand, the experimental
hybrid pumping is often superimposed by a sub-gap leakage that is
too strong to be explained by CPE cotunneling
~\cite{Pekola-NaturePhysics,Kemppinen,Kemppinen2009}, and possibly
caused by structural non-idealities of the sample. This extra
leakage has been described phenomenologically with a model suggested
for the effect of Cooper-pair breaking ~\cite{Dynes}. Here we show
that, specifically to the hybrid devices, already the lowest-order
model predicts accuracy improvement due to implementation of the
resistors. The resistors are therefore expected to suppress a broad
spectrum of unwanted processes of different perturbation orders.
Whereas high-order tunneling usually produces a small current
contribution only, a well-measurable leakage suppression can be an
experimental evidence of the resistors efficiency.

For our experiment, we fabricated hybrid devices of four
different types: SINIS and NISIN type transistors, with and without
chromium resistors (below we denote the devices with resistors as
R-SINIS and R-NISIN). We used a trilayer PMMA/Ge/Copolymer mask and
the shadow evaporation technique \cite{Shadow technique} for the
structure of CrO$_{\text x}$ ($\unit[11]{nm}$ of Cr for the
resistors, evaporated in O$_{\text 2}$), Al ($\unit[18]{nm}$,
oxidized after evaporation), and, finally, $\unit[15]{nm}$ of Cr as
a counter electrode. One of the R-SINIS devices is shown in
Fig.~\ref{Fig1}. The basic results of this work are demonstrated,
using transistor samples SINIS--1,2 and R-SINIS--1, and a reference
single junction SIN--1 with the following parameters (respectively):
total asymptotic resistances $R_{\text N} = \unit[140, 195, 355,]{}$
and $\unit[95]{k\Omega}$ and charging energies of the transistors
$E_{\text C} =\unit[110,140]{}$ and $\sim\unit[200]{\mu eV}$.
The latter estimate is based on scaling the $E_{\text C}$ of the
bare transistors with the tunneling resistance ratios. The resistance
of Cr-lines was $R = \unit[80]{k\Omega} \pm 10\%$ with a small
non-linearity, observed as a zero-bias conductance dip $< 50 \%$
at $T < \unit[100]{mK}$. The effective impedance seen by one of
the identical junctions through the capacitance of the second one is $R/4$.
With our choice of impedance, $R/4 < R_{\rm Q} \equiv h/e^2 \approx
\unit[25.8]{k\Omega}$, the reduction of the single-particle
tunneling rates (and, thus, that of the pumping frequency) is still
moderate~\cite{IngNaz}.

\begin{figure}[t]
\centering%
\includegraphics[width=\columnwidth]{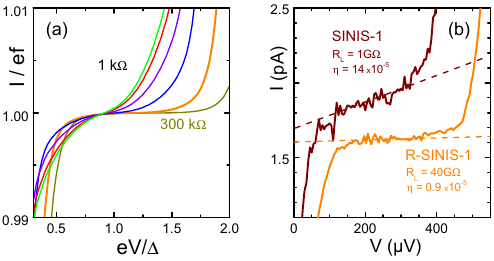}
\caption{ (a) Pumping plateaus calculated for different resistances
$R=\unit[1, 3, 10, 30, 100, 300]{k\Omega}$ and fixed $R_{\text
N}=\unit[200]{k\Omega}$, $E_{\text C}=\Delta=\unit[220]{\mu eV}$,
$T=\unit[100]{mK}$, $f=\unit[10]{MHz}$, $C_{\text g}V_{\text
{g0}}=0.5e$, $C_{\text g}V_{\text {A}}=0.6e$, where $C_{\text g}$ is
a gate capacitance, and the open-state dc-leakage
$I = 5\cdot 10^{-5} \times (R_{\text N})^{-1} V$.
 (b) The experimental plateaus measured under the same
conditions as in (a). The plateau of the SINIS--1 device is shifted
by $+\unit[0.3]{pA}$ for clarity. The dashed lines show the minimum
tilt along the plateau.} \label{PumpLowF}
\end{figure}

We modeled the single-electron tunneling rates in R-SINIS devices
with the $P(E)$ function formalism (see, e.g., Ref.~\cite{IngNaz}),
where $P$ can be interpreted as the probability of the exchange of
energy $E$ between the SINIS transistor and its electromagnetic
environment. The leakage was phenomenologically modelled by smearing
the BCS density of states by introducing an appropriate lifetime of
quasiparticles~\cite{Dynes}, resulting in a linear dc sub-gap
currents, $I = 5\cdot 10^{-5} \times (R_{\text N})^{-1} V$, for the
unblocked bare devices. For low $R = \unit[1]{k\Omega}$, which
corresponds to the resistance of the Cr island itself, the simulated
leakage is almost linear and close to that of the bare transistor.
However for $R = \unit[100]{k\Omega}$, the simulated sub-gap leakage
is clearly nonlinear and it is suppressed by almost an order of
magnitude at $eV=\Delta$.

Figure~\ref{PumpLowF}(a) shows the pumping plateaus at
$\unit[10]{MHz}$, calculated for the broad range of impedances $R$.
We found that the high-ohmic environment extends the plateau and
shifts its inflection point towards higher voltages. The minimum of
the slope is found at the voltages $eV/\Delta\approx 0.9$ and 1.15
for $R = \unit[1]{k\Omega}$ and $R = \unit[100]{k\Omega}$,
respectively. Interestingly, the minimum of the slope appears to be
at the crossing point with $I=ef$, which might help locating the
optimum operating point in practice. The lowest of possible slopes
is expected for $R = \unit[100]{k\Omega}$, being about 17 times
lower than that in the case $R = \unit[1]{k\Omega}$.

The experimental plateaus for the SINIS and R-SINIS devices are
compared in Fig.~\ref{PumpLowF}(b), demonstrating the effect of the
high-ohmic environment on the quantized current plateau at a low
frequency. The figure of merit can be expressed by means of the
leakage parameter $\eta=R_\mathrm{N}/R_\mathrm{L}$, where
$R_\mathrm{L}$ is the lowest slope along the current plateau.
Consistent with the model prediction, the resistor forced a
considerable leakage suppression, corresponding to $\eta^{\mathrm
{(R-SINIS)}}/\eta^{\mathrm {(SINIS)}} \approx 16$. Furthermore, the
plateau of the R-SINIS sample extends to much higher voltages than
has ever been observed for a bare hybrid
turnstile~\cite{Pekola-NaturePhysics,Kemppinen,Kemppinen2009}.

\begin{figure}[b]
\centering%
\includegraphics[width=\columnwidth]{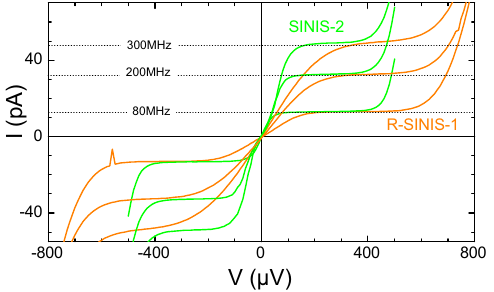}
\caption{Pumping at higher frequencies. The gate sweep was optimized as in Fig.~\ref{PumpLowF}.}
\label{PumpHighF}
\end{figure}

At higher frequencies, a noticeable slow-down of tunneling shows up
for the R-SINIS devices, as compared to the bare ones. In the
pumping $I$--$V$ curves, Fig.~\ref{PumpHighF}, the rise to the
plateau requires higher voltage, and the current versus
gate-amplitude curve (not shown) demonstrates considerable
back-tunneling effects \cite{Kemppinen2009}. At the frequencies
above $\unit[100]{MHz}$, the plateau starts to be deteriorated.
According to our simulations, pumping in this frequency range is
subject to intensive electron heating in the island. We note that,
in the bare SINIS devices, the pumping plateaus were observed up to
the high-frequency roll-off, $f \sim \unit[1]{GHz}$, of our gate
line.

\begin{figure}[t]
\centering%
\includegraphics[width=\columnwidth]{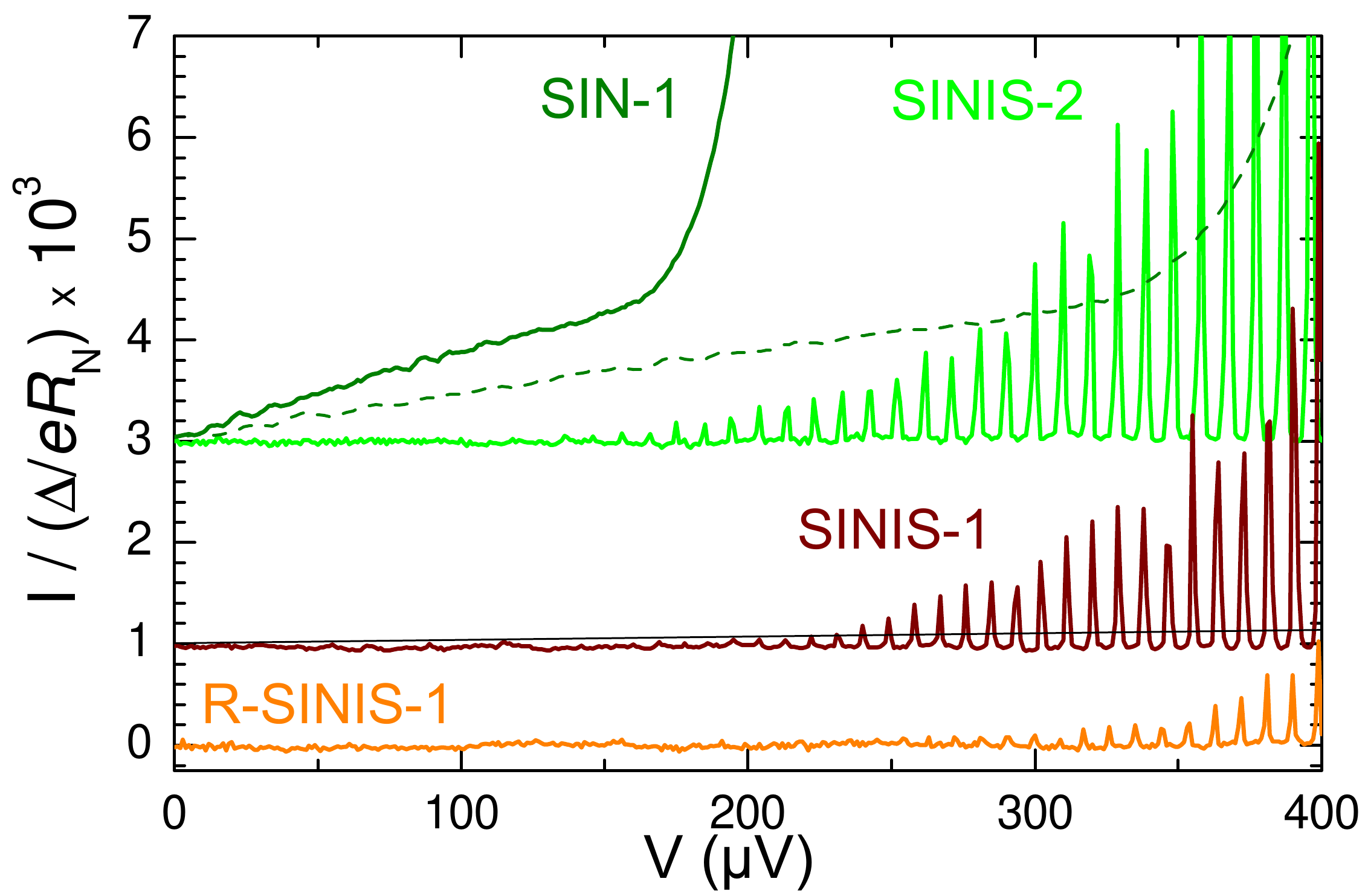}
\caption{Normalized dc $I$--$V$ characteristics. The $I$--$V$
envelopes of the transistors were measured, while simultaneously
sweeping the bias and the gate voltages, the latter over many
oscillation periods. See the text for further details. }
\label{DCLeak}
\end{figure}

For a more detailed insight into the subgap processes, we studied dc
envelopes, Fig.~\ref{DCLeak}, of both SINIS and R-SINIS devices in a
wide range of $R_{\text N}$ and $E_{\text C}$. The high-ohmic
environment is found to dramatically suppress the leakage, whereas
the leakage magnitude appears to be of the same order with the slope
of the pumping plateau. In our simulations, Fig.~\ref{PumpLowF}(a),
we apply a phenomenological leakage slope (the thin solid line in
Fig.~\ref{DCLeak}), which was close to the averaged envelope and
provided us with a correct prediction. However, the detailed
mechanism is probably more complex and includes the processes beyond
the lowest-order model. The dashed line in Fig.~\ref{DCLeak} shows
the expectation for the bare transistor SINIS--2, in the gate-open
state, plotted by appropriate scaling of the $I$--$V$ curve of the
representative junction SIN--1, designed to be a half of the SINIS
layout. The scaling approach is physically relevant, if we assume
that the subgap current appears due to unblocked but still
correlated SET transport through the subgap ( e.g., "poisoning")
states in the superconductor.

In contrast to the expected linear low-bias slope, the experimental
envelopes show a clear voltage threshold, followed by a steep increase
of the subgap current. One plausible interpretation for the peaks is
Andreev reflection (AR), which here may be stronger than in other realizations
\cite{Pekola-NaturePhysics,Kemppinen, Kemppinen2009} because of
lower barrier quality of Al/AlO$_{\text{x}}$/Cr junctions. We note that at
low charging energies, $E_{\text{C}}<\Delta$, AR cycles can be launched in
the following way. First, a nonequilibrium quasiparticle (often generated in the
superconducting leads by external noise or single photons due to imperfect
filtering or shielding) tunnels from a lead to the island and gets trapped
there due to the charge hysteresis induced by $\Delta$. Let us assume, e.g.,
that the gate charge $C_{\text g}V_{\text{g}}/e$ is close to 0 and the island
is trapped to the charge state 1. Now, AR cycles can contribute to the current
by tunneling events between the charge states $\pm 1$. The exact trapping
mechanisms can vary greatly in different experiments, see, e.g.,
Ref.~\cite{Hergenrother} and the citations therein. Experimenting with
several devices, we observed a large spread of leakage thresholds which
may be related to the trapping-induced AR. The effective suppression of
leakage peaks by the resistors indicates thus their efficiency against
higher-order processes. Together with the requirement
$E_{\text{C}} > \Delta$ \cite{PekolaAverin}, a radical improvement can
be expected.

To conclude, we demonstrate the effect of high-ohmic environment on
the hybrid turnstile. In the lower MHz-range, an order of magnitude
improvement of the current plateaus was demonstrated in simulations
and observed in experiment. Further analysis of the effect of the
high-ohmic environment on the higher-order processes is necessary
for developing the hybrid turnstile towards metrological
applications. Also, the effect of cooling or heating of the
island~\cite{SET-cooling} should be studied in more detail for
further understanding of the device frequency limitations.

Fruitful discussions with M.~M\"ott\"onen, O.-P. Saira, M.~Meschke
and V.~Bubanja are gratefully acknowledged. A.K. thanks
V\"ais\"al\"a foundation for financial support. The work was
partially supported by Technology Industries of Finland Centennial
Foundation, the Academy of Finland, and by the European Union through the project SCOPE.

\end{document}